\documentclass{nature}

\def\1023{PSR J1023+0038}
\def\ltsima{$\; \buildrel < \over \sim \;$}
\def\simlt{\lower.5ex\hbox{\ltsima}}
\def\gtsima{$\; \buildrel > \over \sim \;$}
\def\simgt{\lower.5ex\hbox{\gtsima}}

\usepackage{float}
\usepackage{amsmath}
\usepackage{amssymb}
\usepackage{graphicx}
\usepackage{epstopdf}
\usepackage{tabu}
\usepackage{threeparttable}
\usepackage{rotate}
\usepackage{lscape}
\usepackage{longtable}
\usepackage{url}
\usepackage{color}
\usepackage{multibib}

\title{Optical pulsations from a transitional millisecond pulsar}

\author{F. Ambrosino$^{1, *}$, A. Papitto$^{2, *, \dagger}$, L. Stella$^2$, F. Meddi$^1$,  P. Cretaro$^3$, L.~Burderi$^4$, T.~Di Salvo$^5$, G.~L.~Israel$^2$, A. Ghedina$^6$, L. Di Fabrizio$^6$ \& L. Riverol$^6$}
\makeatletter
\let\saved@includegraphics\includegraphics
\renewenvironment*{figure}{\@float{figure}}{\end@float}
\makeatother

\begin{document}

\maketitle

\begin{affiliations}
 \item Dipartimento di Fisica, Universit\`a di Roma ``La Sapienza'', Piazzale Aldo Moro, 5, 00185 Roma, Italy.
 \item INAF - Osservatorio Astronomico di Roma, Via Frascati, 33, 00040 Monte Porzio Catone (RM), Italy.
\item INFN - Sezione di Roma 1, Piazzale Aldo Moro, 5, 00185 Roma, Italy.
\item Dipartimento di Fisica, Universit\`a di Cagliari, SP Monserrato-Sestu, Km 0.7, I-09042 Monserrato, Italy
\item Dipartimento di Fisica e Chimica, Universit\`a di Palermo, via Archirafi 36, 90124 Palermo, Italy
\item Fundaci\'{o}n Galileo Galilei - INAF, Rambla Jos\'{e} Ana Fern\'{a}ndez P\'{e}rez, 7, 38712 Bre\~{n}a Baja, TF, Spain.
  \item[*] These authors contributed equally to this work.
   \item[$\dagger$]  corresponding author (e-mail: alessandro.papitto@oa-roma.inaf.it)
\end{affiliations}

\begin{abstract}

Millisecond pulsars are neutron stars that attain their very fast
rotation during a $10^8$-$10^{9}$~yr-long phase of disk-accretion of
matter from a low mass companion
star\cite{alpar1982,radhakrishnan1982}. They can be detected as
accretion-powered millisecond X-ray pulsars if towards the end of this
phase their magnetic field is strong enough to channel the in-flowing
matter towards their magnetic poles\cite{wijnands1998}. When mass
transfer is reduced or ceases altogether, pulsed emission generated by
magnetospheric particle acceleration and powered by the star rotation
is observed, preferentially in the radio\cite{backer1982} and
gamma-ray\cite{abdo2009} bands. A few transitional millisecond pulsars
that swing between an accretion-powered X-ray pulsar regime and a
rotationally-powered radio pulsar regime in response to variations of
the mass in-flow rate have been recently
identified\cite{archibald2009,papitto2013}.  Here we report the
detection of optical pulsations from a transitional pulsar, the first
ever from a millisecond spinning neutron star.  The pulsations were
observed when the pulsar was surrounded by an accretion disk, and
originated inside the magnetosphere or within a few hundreds of
kilometers from it. Energy arguments rule out reprocessing of
accretion-powered X-ray emission and argue against a process related
to accretion onto the pulsar polar caps; synchrotron emission of
electrons in a rotation-powered pulsar magnetosphere\cite{cocke1969}
seems more likely.

\end{abstract}

{\1023} is a 1.69~ms spinning neutron star in a 4.75~hr orbit around a
0.2~M$_{\odot}$ main-sequence-like companion, located at a
distance\cite{deller2012} of 1.37~kpc. It was discovered in 2008 as a
rotationally-powered radio pulsar\cite{archibald2009} releasing a spin
down power\cite{deller2012} of $4.3\times10^{34}$~erg~s$^{-1}$,
corresponding to a surface magnetic field of $\simeq 10^8$~G.  In June
2013 radio pulsations disappeared and an accretion disk developed,
accompanied by a factor of $\sim10$ increase both in the
average\cite{stappers2014,patruno2014,tendulkar2014} and pulsed X-ray
flux\cite{archibald2015,bogdanov2015} as compared to the radio pulsar
regime.  This increase was ascribed to channeled mass accretion onto
the pulsar polar caps, even though the X-ray
luminosity\cite{papitto2015} ($\simeq 7\times10^{33}$~erg~s$^{-1}$,
0.3-79~keV) was considerably lower than both the luminosity of
accreting millisecond pulsars in outburst, and the luminosity
corresponding to the mass accretion rate required to overcome the
centrifugal barrier of the pulsar rotating magnetosphere and its
propelling effect (assuming matter reaches the neutron star).  Its
gamma-ray\cite{stappers2014} and flat-spectrum continuum radio
emission, consistent with a jet-like outflow\cite{deller2015} add to
the complex phenomenology of {\1023} in the accretion disk state,
possibly being manifestations of the interaction between the disk, the
pulsar magnetosphere and its
wind\cite{stappers2014,takata2014,cotizelati2014,papitto2015}.
Archival observations\cite{archibald2009} indicate that {\1023} was in
a similar state in 2001. The optical counterpart of {\1023}, AY Sex,
in the disk state is a $g\simeq16.7$~mag blue object that emits an
average luminosity\cite{cotizelati2014,shahbaz2015} of
$L_{opt}\simeq10^{33}$~erg~s$^{-1}$ in the 320--900~nm band.  Most of
the optical/UV emission originates from the outer regions of the disk,
and from the companion star's face illuminated by the pulsar
high-energy radiation, which drives a $\Delta g\simeq0.4$~mag
modulation\cite{cotizelati2014} of the optical flux at the orbital
period of the system.

\begin{figure}
 \begin{center}
 \includegraphics[scale=0.9]{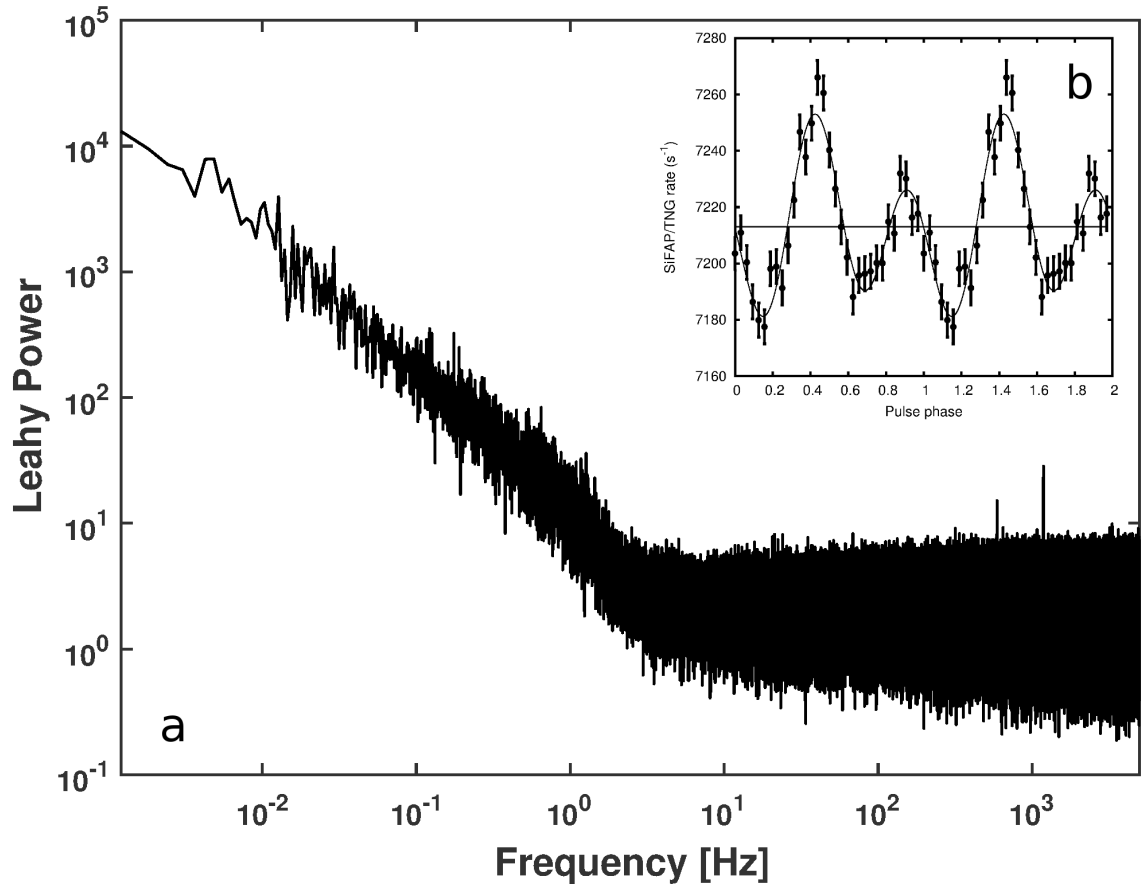}
 \end{center}
 \caption{\footnotesize Coherent optical pulsations of {\1023}.  {\bf a}
   Average Fourier power spectral density of the 320-900~nm optical
   photons observed by SiFAP mounted at the TNG, during four almost
   consecutive observations starting on March 2, 2016 at 21:40 (UTC),
   for an exposure of 13.2~ks. The power spectrum was obtained
   sampling the time series at a time resolution of $0.1$~ms and
   averaging the density measured in eight intervals, each
   $1.65$~ks-long. The time of arrival of optical photons were
   corrected for a known systematic drift of the SiFAP clock (see
   Methods), and converted to the barycenter of the Solar System and
   to the line of nodes of the binary system hosting {\1023}, using
   the parameters listed in Table~\ref{tab:eph}. The peaks at $592.4$
   and $1184.4$~Hz represent the first and the second harmonic of the
   coherent signal detected at the pulsar spin frequency. {\bf b}
   Average, background subtracted pulse profile obtained by folding
   the optical photons detected during the four TNG observations
   around $P_{opt}=1.68798744$~ms (see Table~\ref{tab:eph}). The pulse
   profile is sampled by 32 phase bins, error bars show uncertainties
   at 1-$\sigma$ confidence level, and two cycles are plotted for
   clarity. The dashed solid line is a Fourier decomposition with two
   harmonic components with fractional amplitudes
   $A_1=(0.18\pm0.02)\%$ and $A_2=(0.34\pm0.02)\%$ (giving a total
   amplitude $A=(A_1^2+A_2^2)^{1/2}=(0.38\pm0.03)\%$) with respect to
   the net optical flux of {\1023} averaged over an orbital cycle,
   $K=8185\pm76$~s$^{-1}$. The variance of the profile with respect to
   a constant is $\chi^2=438$ for $31$ degrees of freedom.  The
   probability of observing a pulse profile with such a high variance
   if it were due to a statistical fluctuation was just $p<10^{-65}$,
   allowing us to rule out such a null hypothesis at a high
   confidence.}
 \label{fig:pds_profile}
 \end{figure}

\begin{figure}
\begin{center}
\includegraphics[scale=0.9]{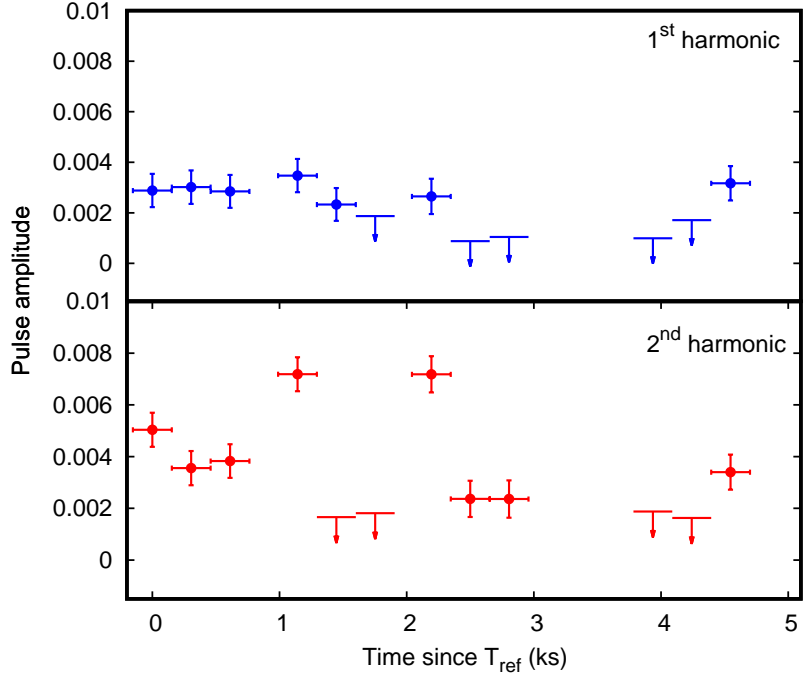}
\end{center}
\caption{\footnotesize Variability of the coherent optical pulsed
  amplitude of {\1023}.  Background subtracted fractional amplitudes
  of the first ($A_1$; panel {\bf a}) and second ($A_2$; panel {\bf b} )
  harmonic of pulse profiles obtained folding the SiFAP time
  series around $P_{opt}=1.68798744$~ms over 1.1~ks-long intervals, sampling the
  period in $n=16$ phase bins. The pulse profiles were modelled with
  the function, $R(t)=K\{1+A_1 sin{[2\pi/P_{opt}(t-\phi_1)]}+A_2
  sin{[4\pi/P_{opt}(t-\phi_2)}\}$, with $K=8185\pm76$~s$^{-1}$, equal
    to the average net optical flux of {\1023}, measured by modeling
    the counting rates observed during the four observations (spanning
    4.85~hr) with a sine function at the orbital period of the binary
    system, $4.75$~hr (see Table~\ref{tab:eph}). Error bars show
    uncertainties at 1-$\sigma$ confidence level, upper limits were
    evaluated at the 3-$\sigma$ confidence level.}
\label{fig:pulseamplitude}
\end{figure}

During a 4 hour-long observation carried out on 2016, March 2-3 with
SiFAP, a fast photometer with 25~ns time resolution that was mounted
at the 3.58~m INAF's Telescopio Nazionale Galileo (TNG) in La Palma,
Spain (see Methods), we discovered optical (320-900~nm) pulsations at
the spin period of {\1023} (see Fig.~\ref{fig:pds_profile}).  In order
to detect the signal we corrected the arrival times of the optical
photons for the light travel time delays introduced by the pulsar
orbit by using the ephemeris derived from the X-ray
pulsations\cite{jaodand2016} (see Tab.~\ref{tab:eph}). {\1023} was in
a disk dominated state at the time of the SiFAP measurement, as
inferred from Swift X-ray Telescope (XRT) observations performed
within a few days from the optical observations (see Methods).  The
optical pulse profile comprises two peaks with a fractional amplitude
that varied significantly over $20$ minutes-long time intervals
between a maximum value of $A=(0.80\pm0.07)\%$ of the total average
emission, and values below the detectability level ($A<0.19\%$; see
Fig.~\ref{fig:pulseamplitude}).  The maximum observed pulsed flux
corresponds to $L_{pulsed}\approx 0.01L_{opt} \approx
10^{31}$~erg~s$^{-1}$. The lack of simultaneous X-ray observations
prevented us from searching for possible changes taking place at the
time-scales set by the variability of the X-ray pulse
amplitude\cite{archibald2015}.

The region responsible for the optical pulsations cannot be larger
than $c\ P_{spin}\sim 500$~km, as light propagation delay would smear
them out. Comparison of the projected semi-major axis $a\sin i $ and
epoch of the pulsar passage at the ascending node $T_{asc}$ derived
from the orbital modulation of the optical pulse period with the
ephemeris derived from X-ray pulsations, showed that the region
emitting the optical pulsation must be centered within $\delta
r_{\rho}\sim\sigma_a \simeq 30 / \sin(i)$~km and $\delta r_\theta\sim
2\pi a \sigma_{T_{asc}}/P_{orb} \simeq300/ \sin(i)$~km of radial and
azimuthal distance from the X-ray pulsar, respectively (see
Table~\ref{tab:eph} and Methods). These values are comparable to or
somewhat larger than than both the $r_{cor}\sim 24$~km corotation
radius (for a 1.4~M$_\odot$ neutron star) of {\1023}, {\it i.e} the
radius where the angular velocity of the pulsar magnetosphere equals
the Keplerian velocity, and the $r_{lc}\sim 80$~km light cylinder
radius where closed magnetic field lines travel at the speed of light.

Reprocessing of the X-ray pulsations at the surface of the companion
star and/or in the outer disk region, as observed in some X-ray
binaries hosting a strongly magnetized and slowly rotating accreting
pulsar\cite{davidsen1972}, is ruled out as the origin of the optical
pulsations from {\1023}; the reprocessing regions would have a very
different light travel time-delay orbital signature than that of a
region $\delta r\simeq 30-300 / \sin(i)$~km away from the neutron
star, and their size would greatly exceed the maximum beyond which
pulsations are washed out ($cP_{spin}\sim 500$~km). The above problems
could be circumvented if the X-ray pulsations were reprocessed in the
innermost regions of the disk close to the boundary with the pulsar
magnetosphere at $r_{cor}$. However, reprocessing of the X-ray
luminosity of {\1023} ($L_X=7\times10^{33}$~erg~s$^{-1}$) by an area
of $\sim \pi r_{cor}^2$ located at a distance $r_{cor}$ from the
pulsar, would convert to a brightness temperature of
$T_{irr}=(L_X/4\pi\sigma r_{cor}^2)\simeq 1.1\times10^{6}$~K. The
optical output would be more than five thousands times smaller than
the observed pulsed flux, ruling out also this interpretation.

If X-ray pulsations observed in the disk state are due to channeled
accretion onto the magnetic polar regions of the neutron
star\cite{archibald2015,papitto2015b}, one may wonder whether the same
polar hotspots could give rise to the the observed optical pulses.
The X-ray spectrum of accreting millisecond pulsars is modeled with
unsaturated Comptonisation of soft 0.5-1~keV photons emitted from the
polar hotspots, by thermal electrons with temperature of tens of keV
presumably located in the accretion
column\cite{gierlinski2002}. Cyclotron emission by the same electrons
in the $B_s \simeq 10^8$~G magnetic field of this pulsar takes place
at $E_{cyc}\simeq 1\,(B_s/10^8\,\mbox{G})$~eV energies.  Optical
pulsed flux might result from self-absorbed cyclotron emission in the
optically-thick regime, with a Rayleigh-Jeans spectrum at the
temperature of the Comptonising electrons extending over a range of
cyclotron harmonics\cite{frank2002}.  We adopt for {\1023} an electron
temperature $k T_{el}$ of $100$~keV (photons up to $\sim 80$~ keV have
been detected from it\cite{tendulkar2014}) and an accreting polar
region of conservatively large area $A\simeq \pi R_*^2
(R_*/R_{cor})\simeq 100$~km$^2$ (ref. \cite{frank2002,Romanova2004}),
where $R_*=10$~km is the neutron star radius. The maximum pulsed
luminosity in the visible band is
\begin{equation}
  \label{eq:cyc}
L_{cyc} \simeq 3\times10^{29} \, \left(\frac{A}{100\,\mbox{km}^2}\right) \, \left(\frac{kT_{el}}{100\,\mbox{keV}}\right)\,\mbox{erg s}^{-1},
\end{equation}
 more than $\sim30$ times lower than the observed pulsed luminosity,
 $L_{pulsed}\approx10^{31}$~erg~s$^{-1}$. This does not favour an
 interpretation of the optical pulses in terms of cyclotron emission
 from electrons in the accretion columns; a detailed modelling will
 assess whether a larger optical efficiency can be achieved.

\begin{figure}
  \centering
  \includegraphics[width=13cm]{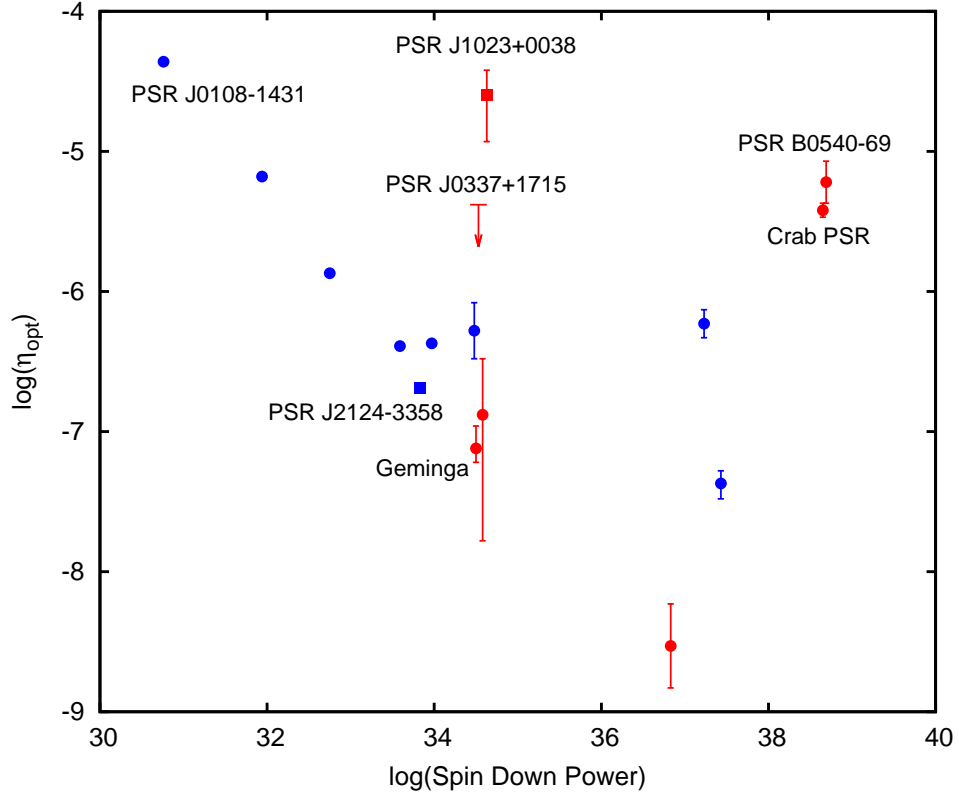}
  \caption{ \footnotesize { Optical efficiency of rotation-powered
      pulsars, $\eta_{opt}=L_{opt}/\dot{E}_{sd}$ as a function of the
      pulsar spin down power, $\dot{E}_{sd}$. Red symbols indicate
      optical pulsars or sources for which an upper limit was set on
      the pulse amplitude. Blue symbols mark pulsar optical
      counterparts proposed on the basis of a detection of an
      (unpulsed) optical source at a position compatible with the
      pulsar, under the assumption that the optical output traces
      magnetospheric emission. Circles refer to high-magnetic field
      ($B_s>10^{-11}$~G), slow ($P>0.1$~s) pulsars, squares are used
      to mark low-magnetic field ($B_s<10^9$~G) millisecond
      ($P<0.01$~s) pulsars. The optical luminosity was evaluated in
      the B-band, except for a few cases (see Supplementary
      Table~1). Error bars show uncertainties at 1-$\sigma$ confidence
      level. The mean pulsed flux of {\1023} was evaluated as
      $<A>=(0.38\pm0.03)$ per cent of the average de-reddened B-band
      magnitude\cite{bogdanov2015}, $16.77\pm0.05$.  Values of the
      pulsar distance and spin down power were taken from ATNF pulsar
      catalog\cite{manchester2005}.}}
\label{fig:Edot}
\end{figure}

Finally we explore the rotation-powered regime of pulsars.
Synchrotron radiation from secondary relativistic electrons and
positrons in the magnetosphere of pulsars is generally believed to
produce non-thermal pulsed emission from the optical to the X-ray
band\cite{pacini1983,romani1996}.  Pulsed emission in the visible band
has been detected from five rotation-powered
pulsars\cite{cocke1969,mignani2011}. They are all isolated,
high-magnetic field ($>10^{12}$~G) pulsars with young to moderate ages
of $10^3 - 10^5$~yr.  Their efficiency in converting spin down power
to optical pulsed luminosity spans a broad range from $\eta_{opt}\sim
5\times 10^{-6}$ (e.g., the Crab pulsar) to $\sim 2 \times10^{-9}$
(see red circles in Fig.~\ref{fig:Edot} and Supplementary
Table~\ref{tab:S1}). The efficiency of {\1023}, $\eta_{opt} \sim 2\times10^{-5}$,
is higher; note that two middle-aged pulsars with a spin down power
comparable to that of {\1023} (e.g. Geminga) have $\eta_{opt}\simlt
10^{-7}$.  A few more pulsar optical counterparts were proposed based
on the positional coincidence of an optical source (blue circles in
Fig.~\ref{fig:Edot}, see Supplementary Table~1); an increasingly
higher efficiency is observed for system with lower and lower spin
down power and ages exceeding $10^6$~yr.

Comparison of {\1023} -- the first optical millisecond pulsar ever
detected -- with other rotation-powered millisecond pulsars is
necessarily limited.  A recent search\cite{strader2016} did not detect
optical pulses down to a magnitude of $g\sim25$ from PSR~J0337+1715, a
millisecond radio pulsar of similar spin down power and distance as
{\1023}, whose average optical pulses correspond to $g\simeq22.5$.  A
candidate optical counterpart of PSR~J2124-3358, a close-by 4.4~ms
binary radio pulsar with spin-down power $\sim 6$ times lower than
that of {\1023}, has been recently reported\cite{rangelov2017}; its
optical luminosity of $\sim 10^{27}$~erg~s$^{-1}$ gives
$\eta_{opt}\sim 10^{-7}$.  Therefore the optical efficiency of {\1023}
appears to be orders of magnitude higher than that of other
millisecond pulsars, and lower only than the efficiency of the
proposed optical counterpart\cite{mignani2008} of the old
($1.7\times10^8$~yr), isolated 0.8~s pulsar, PSR~J0108-1431.

The feature that singles out {\1023} among rotation-powered pulsars is
the presence of an accretion disk.  If the disk does not prevent
rotation-powered pulsar mechanisms from working, the interaction
between pulsar magnetosphere\cite{papitto2014,papitto2015} and/or its
wind\cite{stappers2014,takata2014,cotizelati2014} with the disk plasma
may evaporate enough material to disperse radio pulsations and make
them unobserved. Moreover, it would form a shock where electrons and
positrons could be accelerated to the energies ($1-20$ MeV) required
to radiate optical synchrotron photons, in the field of $B\approx
2\times10^{5}$-$10^6$~G of the magnetospheric region inward of the
light cylinder of {\1023}. This might be the reason for the higher
$\eta_{opt}$ of {\1023}. The absence of a dramatic change of the
spin-down rate when the source transitioned from a radio pulsar to a
disk-dominated state\cite{jaodand2016} might provide additional
evidence that a rotation-powered mechanism is working in {\1023},
despite the presence of a disk.

Our discovery of optical pulsations from {\1023} demonstrates that the
magnetosphere of old, weakly magnetic and quickly spinning neutron
stars can give rise to such signals when surrounded by an accretion
disk. This new observational window provides a promising diagnostic to
probe the physics of millisecond pulsars in close binary systems, and
discover new millisecond pulsars in low-mass X-ray binaries and in
unidentified gamma-ray sources obscured in the radio band by matter
enshrouding the system.

The optical pulsations observed from {\1023} are still open to
different interpretations. The pulsed optical luminosity rules out
disk reprocessing of accretion-powered X-ray emission and is also hard
to reconcile with cyclotron emission from matter accreting onto the
neutron star polar caps. We argued that the observed optical pulsed
emission could be due to synchrotron emission by relativistic
electrons in the magnetosphere of a rotation-powered pulsar. We note
that the observed optical efficiency of {\1023} exceeds by about three
decades that of pulsars of comparable spin-down luminosity. That may
be due to the shocked magnetised environment that results from the
interaction between the rotation-powered pulsar magnetosphere and the
accretion disk, a characteristic of {\1023} and likewise other
transitional millisecond pulsars. Future observations will assess
whether the process responsible for optical pulsations from {\1023}
coexist or alternate with accretion-powered X-ray pulsations.

\clearpage
\subsection{Reference list.}
\bibliographystyle{naturemag}

\clearpage

\begin{addendum}
 \item[Correspondence and request for materials]should be addressed to
   A.~P. (e-mail: alessandro.papitto@oa-roma.inaf.it).
\item[Acknowledgements] F.~A., P.~C. and F.~M.  acknowledge the research
   project Protocol \textit{C26A15YCJ4} funded by the Department of
   Physics of the University of Rome "La Sapienza'', for the financial
   support.  A.~P. acknowledges funding from the EU’s Horizon 2020
   Framework Programme for Research and Innovation under the Marie
   Sk{\l}odowska-Curie Individual Fellowship grant agreement
   660657-TMSP-H2020-MSCA-IF-2014, and wishes to thank D.~de Martino,
   N.~Rea and D.~F.~Torres for very useful discussions. L.~B.,
   T.~D.S. and A.~P. acknowledge fruitful discussion with the
   international team on ``The disk-magnetosphere interaction around
   transitional millisecond pulsars'' at ISSI (International Space
   Science Institute), Bern.  L.~B., T.~D.S., G.L.~I. and
   L.~S. acknowledge financial contribution from the agreement ASI-INAF
   I/037/12/0. Results obtained and presented in this paper are based
   on observations made with the Italian Telescopio Nazionale Galileo
   (TNG) operated on the island of La Palma by the Fundaci\'{o}n
   Galileo Galilei of the INAF (Istituto Nazionale di Astrofisica) at
   the Spanish Observatorio del Roque de los Muchachos of the
   Instituto de Astrofisica de Canarias.  The authors gratefully
   acknowledge time Discretionary Observing Time granted by the TNG
   Director, Dr. E.~Molinari and thank Prof. C.~Rossi, for her
   scientific support and contribution.
 \item[Author contributions] F.A. and A.P. jointly coordinated and
   contributed with equal proportion to this study.  F.M, F.A. and
   P.C. conceived and designed the optical photometer. F.A., A.G.,
   L.D.F. and L.R. performed the optical observations. A.P and
   F.A. analysed the data. A.P. and L.S. wrote the paper. A.P., L.S.,
   T.D.S. and L.B. interpreted the results. All authors read,
   commented on and approved submission of this article.
 \item[Competing Interests] The authors declare that they have no
competing financial interests.
\end{addendum}

\clearpage

\clearpage

\clearpage

\clearpage

\begin{table}
\centering
\begin{threeparttable}[h!]
\caption{X-ray\cite{jaodand2016} and optical (this work) ephemerides
  of {\1023}.}
 \begin{tabular}{lcc} 
 \hline
Parameter & X-rays\cite{jaodand2016} & Optical (this work) \\
   \hline
 Right Ascension{\tnote{(a)}}\hspace{0.3cm}, $\alpha$ [J2000] & $10$:$23$:$47.687198$ & ... \\
 Declination\tnote{(a)}\hspace{0.3cm}, $\delta$ [J2000]& $+00$:$38$:$40.84551$  & ...\\
 Reference Epoch, $T_{ref}$ (MJD) & $57449.9028346$ & ... \\
 Spin Period\tnote{(b)}\hspace{0.3cm},  $P_s$ (ms) & $1.68798744420(13)$ & $1.68798744(6)$ \\
 Orbital Period, $P_{orb}$ (s) & $17115.5216592$ & $17116.1 \pm 5.5$ \\
 Projected semi-major axis, $a\sin i/c$ (lt-s) & $0.343356(3)$ & $0.3434(1)$  \\
 Epoch of ascending node passage\tnote{(c)}\hspace{0.3cm},  $T_{asc}$ (MJD)  & $57449.7258(3)$ & $57449.72579(9)$\\
 \hline
 \end{tabular}
 \begin{tablenotes}
   \item[(a)] \footnotesize Astrometric position determined from radio interferometry\cite{deller2012}.
 \item[(b)]\footnotesize Extrapolation to the reference epoch $T_{ref}$
   of the period of the X-ray pulse\cite{jaodand2016} estimated at
   {$T_J=56458$~MJD}, $P_0=1.68798744494252(13)$~s, taking into account
   the spin period derivative,
   $\dot{P}=(8.665+/-0.026)\times10^{-21}$.
 \item[(c)]\footnotesize Extrapolation to the reference epoch
   $T_{ref}$ of the epoch of passage at the ascending node determined
   from the formal orbital solution that models the epochs of passage
   at the ascending node determined from X-ray
   pulsations\cite{jaodand2016}, $T_{asc}=54905.96943473$~MJD,
   $P_{orb}=17115.5502801$~s, $\dot{P}_{orb}=(-1.65\pm
   0.19)\times10^{-10}$.
  
    \end{tablenotes}
 \label{tab:eph}
\end{threeparttable}
\end{table}

\clearpage

{\huge \bf Methods}

\subsection{The optical data set.}
We observed {\1023} with the \textit{Silicon Fast Astronomical
  Photometer}\cite{meddi2012,ambrosino2016} (SiFAP) mounted at the
Nasmyth B focus of the $3.58$~m \textit{Telescopio Nazionale
  Galileo}\cite{barbieri1994} (TNG), in La Palma Spain, exploiting the
focal plane of the \textit{Device Optimized for the LOw RESolution}
(DOLORES) instrument\cite{zappacosta2005}. SiFAP is a 2-channel ultra
fast optical photometer developed at the Department of Physics of the
University of Rome ``La Sapienza''.  It comprises two \textit{Multi
  Pixel Photon Counters} (MPPCs) modules manufactured by Hamamatsu
Photonics, one aimed at measuring photon counting rates from the
target and the other at monitoring a reference star in the
\textit{Field of View} (FoV).  These MPPCs integrate signals in
configurable time windows from $100$~ms down to $1$~ms via a standard
USB interface. They also provide an analog output which can tag the
\textit{Time of Arrival} (ToA) of individual photons with a time
resolution of $25$~ns and a discriminated output capable of counting
photons in time bins of $20$~$\mu$s, through two independent custom
electronic chains. A \textit{Global Positioning System} (GPS) unit
yields a reference time marker via the \textit{Pulse Per Second} (PPS)
signal with $25$~ns resolution at 50\% of the rising edge of the pulse
itself.

Four observations of {\1023} were performed starting on 2016, March 2
at 21:40 (UT), each lasting 3.3~ks (see Supplementary Table~2). A
white filter covering the 320-900~nm band was used in all the
observations (see Supplementary Fig.~1). The airmass ranged between
$1.13$ and $1.74$, while seeing conditions varied from $0.8$~arcsec up
to $3$~arcsec. The FoV of the sensor is $\sim 50$~arcsec$^2$ ensuring
that signal loss was negligible. A reference star, UCAC4~454-048424
($\alpha=10$:$23$:$48.935$, $\delta=+00$:$44$:$52.814$,
$B=16.309$~mag) was also simultaneously observed, at a lower time
resolution ($1$~ms). We estimated the sum of the sky background and of
the dark count-rate by observing a region located $25$~arcsec away
from the target towards the East direction for $120$~s (between the
third and the fourth exposure), which yielded an average countrate of
$B=8559$~s$^{-1}$, amounting to roughly $\simeq50\%$ of the total
photons recorded when pointing at the direction of {\1023}. The
contribution of the dark count rate alone was
$\sim2.5\times10^3$~s$^{-1}$.

\subsection{Temporal Analysis.}
We considered data taken by the SiFAP at the maximum possible time
resolution of $25$~ns. A systematic effect introduced a significant
difference between the actual photon arrival times and those measured
by the SiFAP system quartz clock.  In each of the four observations,
the total time elapsed between the two GPS-PPS signals marking the
beginning and the end of each exposure, which we took as a reference,
$\Delta t_{obs}^{GPS}=3300$~s, was slightly longer than the time
interval measured by the SiFAP clock, $\Delta t_{obs}^{SiFAP}=\Delta
t_{obs}^{GPS}+\delta t$, with $\delta t\simeq6\times10^{-3}$~s. We
assumed that this difference was due to a constant drift of the time
recorded by the SiFAP clock with respect to the actual time, yielding
a cumulative linear effect on the recorded times of arrival,
$t_{SiFAP}$.  We corrected the arrival times by using the following
relation, $t_{corr}=t_{SiFAP}\times (\Delta t_{obs}^{GPS}/\Delta
t_{obs}^{SiFAP})$. We checked this procedure by applying it to the
arrival times recorded during an observation of the Crab pulsar
performed with the $1.52$~m Cassini telescope
({\url{http://davide2.bo.astro.it/loiano/152cm-telescope/}) of the
  Bologna Astronomical Observatory on 2016, December 2 at 00:56 (UT),
  with the same equipment\cite{ambrosino2016}. Using the uncorrected
  arrival times, a spin period of $P_{Crab}^{SiFAP}=33.7296127(23)$~ms
  was obtained; by applying the correction above we recovered a period
  of $P_{Crab}^{corr}=33.7295827(23)$~ms, fully compatible with that
  extrapolated from the monthly radio ephemeris provided by the
  \textit{Jodrell Bank Observatory}
  (\url{http://www.jb.man.ac.uk/pulsar/crab/crab2.txt}),
  $P_{Crab}^{JBO}=33.7295810(11)$~ms. This proved the robustness of
  the SiFAP clock correction. We also determined through laboratory
  tests (using a frequency/period counter) the maximum jitter of the
  system quartz clock period ($P^{clk}=8.000009873(1)\times10^{-9}$~s)
  determined by a thermal drift, $\Delta P_{max}^{clk} =
  P_{max}^{clk}-P_{min}^{clk}=4.8\times 10^{-15}$~s.  The relative
  uncertainty on the measured period introduced by the thermal drift
  is thus $6\times 10^{-7}$, i.e. about sixty times smaller than the
  relative accuracy of our determination of the spin period of {\1023}
  (see Table~\ref{tab:eph}). The effect of thermal drift could be
  safely neglected.

We reported the times of arrival of the corrected photons to the Solar
System Barycentre, using the JPL DE431 ephemeris
(\url{https://ssd.jpl.nasa.gov/horizons.cgi}) and the source radio
astrometric position\cite{deller2015}.  A search for periodicities at
the known spin period of the pulsar (see Table~\ref{tab:eph}) was then
conducted via the epoch folding search
technique\cite{leahy1983,leahy1987}. No signal was detected by
performing an epoch folding search with $n=16$ phase bins on the four
time series not corrected for the pulsar orbital motion, down to an
upper limit on the amplitude of $A<10^{-6}$ (3-$\sigma$ confidence
level). On the other hand, correcting the times of arrivals of optical
photons for the R{\"o}mer delay caused by the pulsar orbital motion
with the ephemerides derived from the X-ray
pulsations\cite{jaodand2016}, allowed us to detect significantly a
coherent signal at the pulsar spin period over time intervals as short
as $1.1$~ks.  The $\chi^2$ of an epoch folding search over a range of
periods around the known spin period of the pulsar is plotted in
Supplementary Fig.~2. The background subtracted pulse profile was
modeled with two Fourier components of fractional amplitude $A_1$ and
$A_2$, correponding to the fundamental and first overtone of the
signal (see the inset of Fig.~1 of the main body), $R(t)=K\{1+A_1
sin{[2\pi/P_{spin}(t-\phi_1)]}+A_2 sin{[4\pi/P_{spin}(t-\phi_2)}\}$.
  The mean optical flux of {\1023} over an orbital cycle, $K=8185 \pm
  76$~s$^{-1}$, was determined by modelling with a sine function at
  the orbital period of the binary system the count-rate observed
  during the four observations, and binned in time intervals of 50~s.
  The fractional amplitudes of these two components varied over the
  1.1~ks intervals considered, from $A_1^{max}=(0.35\pm0.07)\%$ and
  $A_2^{max}=(0.72\pm0.07)\%$, to the non-detection (see
  Fig.~\ref{fig:pulseamplitude}).  The total pulse amplitude was
  evaluated as $A=(A_1^2+A_2^2)^{1/2}$, and varied around an average
  value of $<A>=0.38\%$, with a maximum of
  $A^{max}=(0.80\pm0.07)\%$. The most stringent upper limit on the
  total pulse amplitude of $A<0.19\%$ was obtained by folding the
  first 2.2~ks of the fourth observation.

In order to determine the orbital parameters of the source of the
optical photons, we modeled the temporal evolution of the phase of the
first harmonic component in terms of the difference between the
orbital and spin parameters used to correct the photon arrival times,
and the actual ones\cite{deeter1981}. The procedure was iterated until
no significant corrections to the parameters were found, to within the
uncertainties\cite{papitto2011}. The spin and orbital parameters we
obtained are compatible with those expected from the extrapolation of
the ephemeris measured from the X-ray pulsations to the epoch of
optical observations (see Tab.~\ref{tab:eph}). The uncertainties on
the projected semi-major axis and on the epoch of passage of the
source of optical pulses at the ascending node of its orbit allowed us
to identify the source of optical photons from within $\delta r\simeq
($30--300$) / \sin(i)$~km from the pulsar. The decrease of the signal
power as soon as the values of these two parameters are varied from
their best-fitting values are shown in Supplementary Fig.~3 and
Fig.~4, respectively. This proved that optical pulses came from the
same location of the source of X-ray pulsations.

\subsection{X-ray observations.}
The field around {\1023} was observed twice by {\it
  Swift}\cite{gehrels2004} (see Supplementary Table~2) within a few
days from the TNG observations presented here. Before (seq.~112) and
after (seq.~113) the TNG observations, the source was detected by the
X-ray Telescope\cite{burrows2005} (XRT) with an unabsorbed 0.3-10~keV
X-ray flux of $0.95^{+0.23}_{-0.25}\times10^{-11}$ and
$1.4^{+0.5}_{-0.4}\times10^{-11}$~erg~cm$^{-2}$~s$^{-1}$,
respectively; the X-ray spectrum was described by a power law with
photon index $\alpha=1.7^{+0.5}_{-0.2}$ and
$\alpha=1.6^{+0.7}_{-0.4}$.  The UltraViolet and Optical
Telescope\cite{roming2005} (UVOT) on-board {\it Swift} observed the
field with the UV filter M2 (centered at $\lambda=224.6$~nm). A
counterpart of {\1023} was detected at Vega magnitudes of 16.32(7) and
16.37(8), respectively. Both the observed X-ray and UV fluxes are
typical of the accretion disk state of {\1023}\cite{patruno2014},
allowing us to safely conclude that the source was in such a state
during the TNG observations performed less than a day before the
second Swift observation considered here (id.~113).

\subsection{Data availability.} The barycentered SiFAP data that support the findings of this study are available in the repository \url{figshare} with the identifier \url{doi:10.6084/m9.figshare.5341192}.

\begin{figure}
\begin{center}
\includegraphics[scale=1]{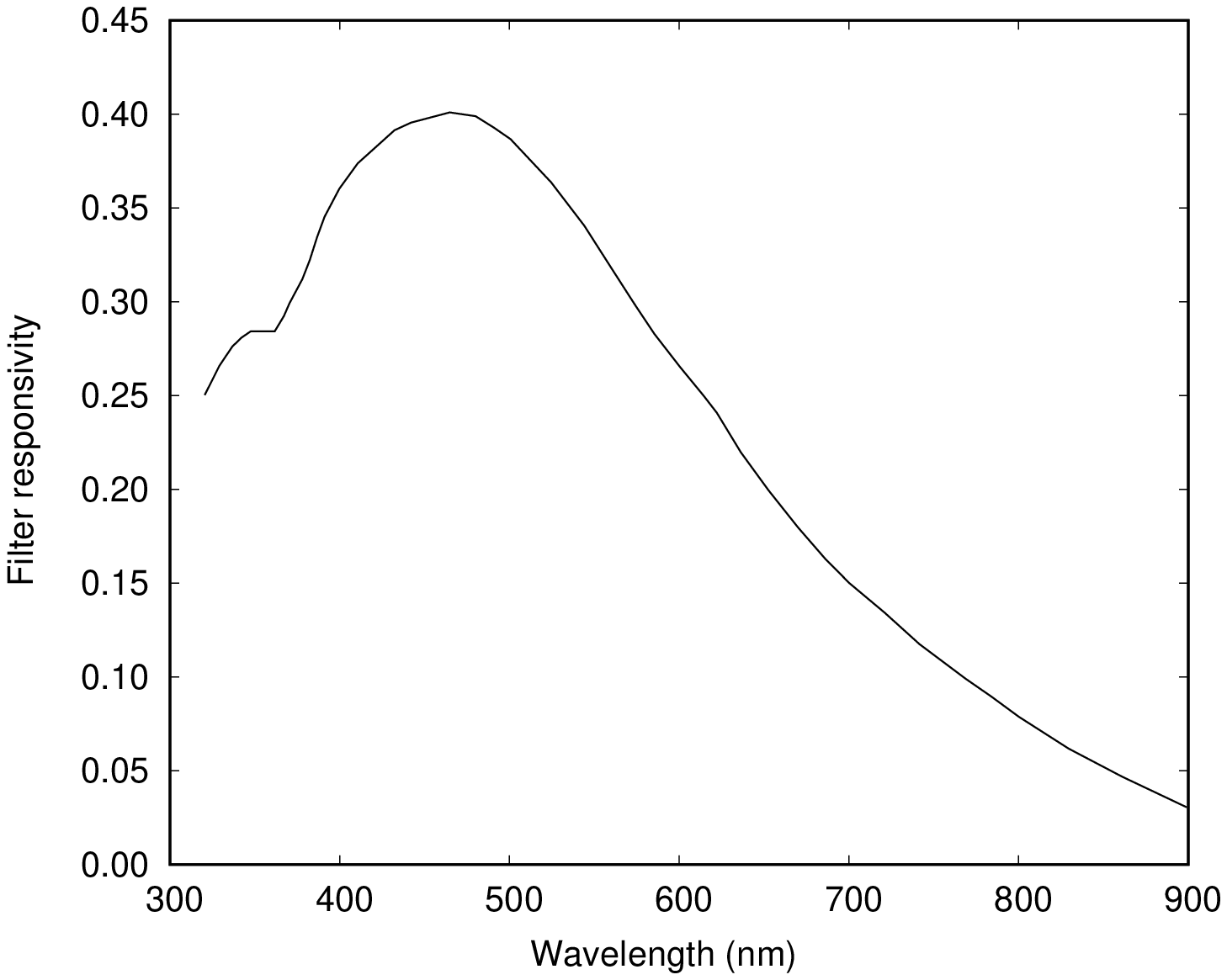}
\end{center}
\renewcommand{\figurename}{Supplementary Figure}
\caption{\footnotesize  Response of the white filter used for all the
  observations as a function of the wavelength. The distribution
  reaches a peak value of $\sim$ 40\% at a wavelength of $450$~nm.}
\label{fig:S1}
\end{figure}

\begin{figure}
\begin{center}
\includegraphics[scale=1]{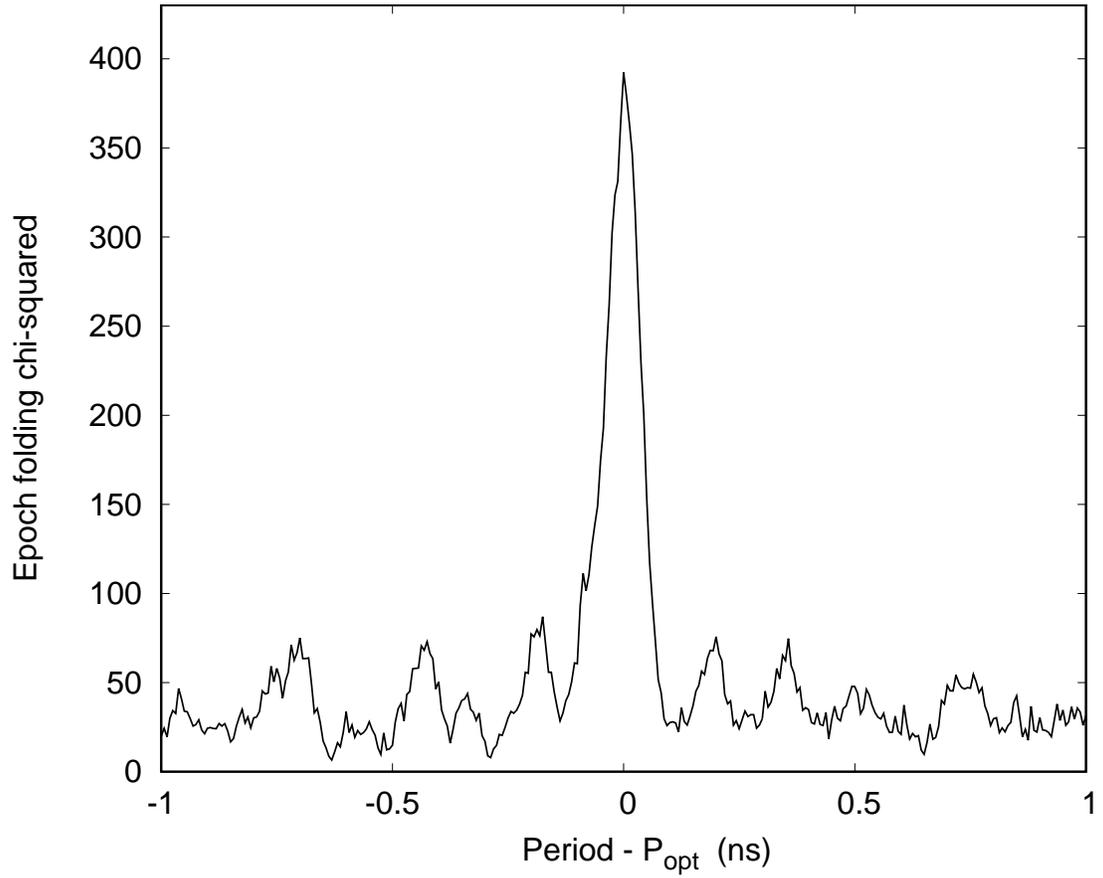}
\end{center}
\renewcommand{\figurename}{Supplementary Figure}
\caption{\footnotesize Chi-squared versus spin period the epoch
  folding search performed over the four TNG observation, centered at
  the best optical period, $P_{opt}=1.68798744$~ms (see Tab.~1 of the
  main manuscript), and obtained using $n=16$ phase bins and a period
  resolution of $5\times10^{-12}$~s.}
\label{fig:S2}
\end{figure}

\begin{figure}
\begin{center}
\includegraphics[scale=1]{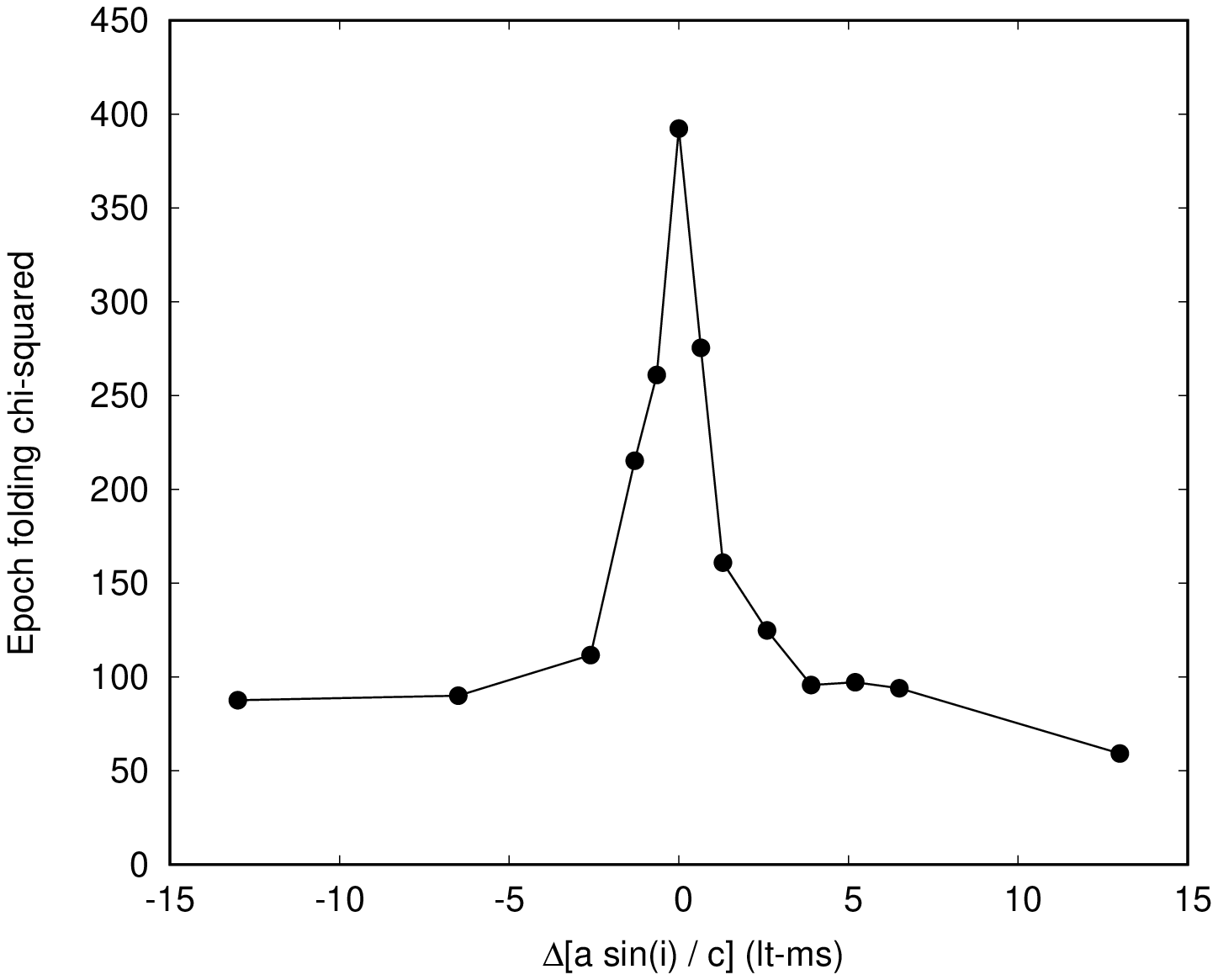}
\end{center}
\renewcommand{\figurename}{Supplementary Figure}
\caption{\footnotesize Chi-squared of the signal obtained from an epoch folding
  search around $P_{opt}$ of the four TNG observations corrected with
  values of the projected semi-major axis differing by $\Delta[a\sin(i)/c]$ from the best-fitting value (see Tab.~1 of the
  main manuscript).}
\label{fig:S3}
\end{figure}       

 \begin{figure}
 \begin{center}
 \includegraphics[scale=1]{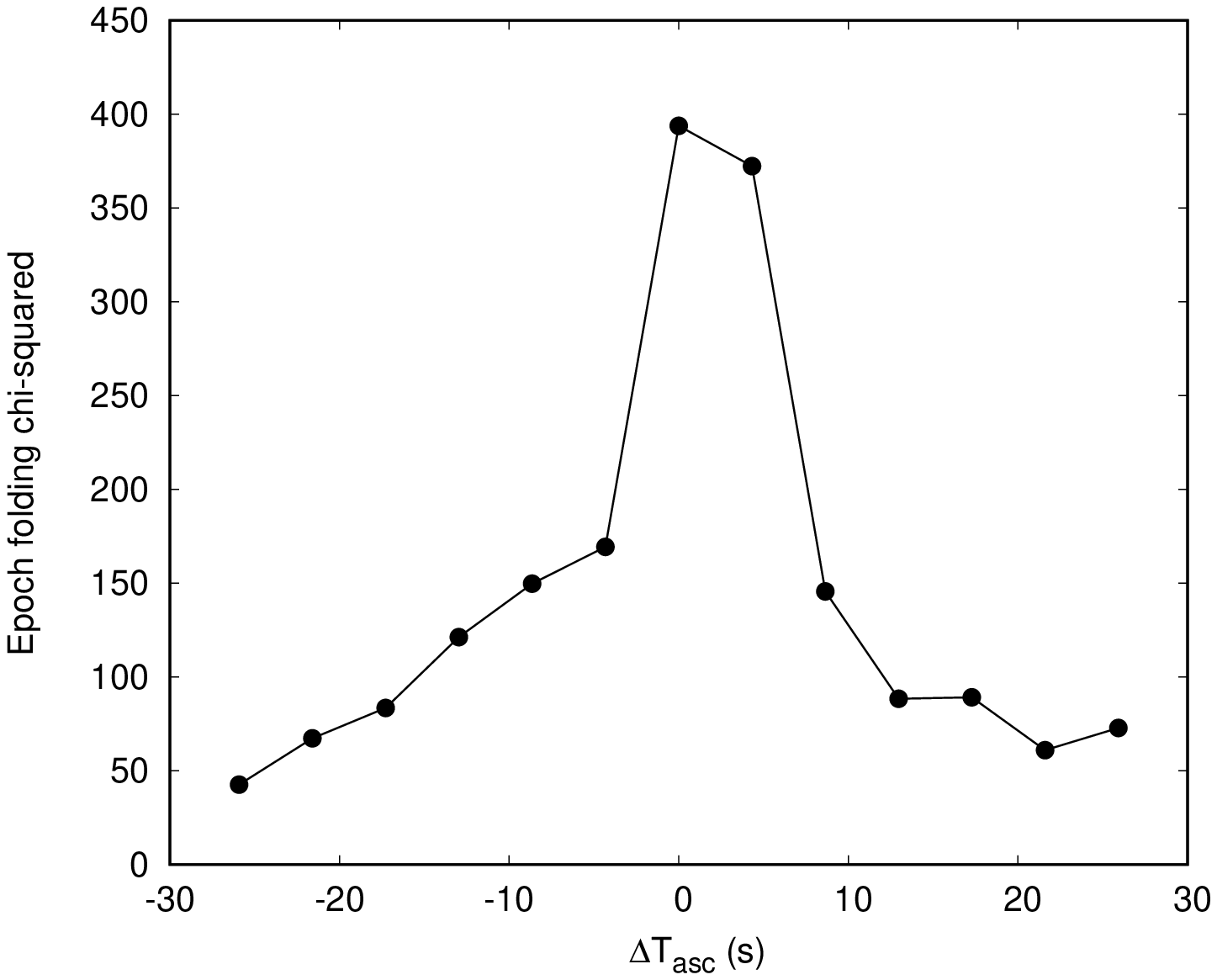}
 \end{center}
 \renewcommand{\figurename}{Supplementary Figure}
 \caption{\footnotesize Chi-squared of the signal obtained from an epoch folding
  search around $P_{opt}$ of the four TNG observations corrected with
  values of the projected semi-major axis differing by $\Delta[T_{asc}]$ 
  from the best-fitting value (see Tab.~1 of the
  main manuscript).
 }
 \label{fig:S4}
 \end{figure}

\clearpage

\begin{table}
  \footnotesize
\centering
\begin{threeparttable}[h!]
\renewcommand{\tablename}{Supplementary Table}
  \caption{Optical luminosity of rotation-powered pulsars.
}
\begin{tabular}{lccccll}
\hline
Pulsar name & $P_{spin}$ $^{[a]}$ & $\log{\tau}$ $^{[a]}$ & d $^{[a]}$ & $\log{\dot{E}}$ $^{[a]}$ & B $^{[b]}$ & $\log{L_{opt}}$ \\
   PSR    &     s     &    yr       & kpc &   erg s$^{-1}$ & mag & erg s$^{-1}$ \\
\hline
\multicolumn{7}{c}{\textbf{Optical Pulsars}} \\
  \hline
B0531+21 (Crab)  & $0.033$   & $3.1$  & $2.0$ &  $38.65$  & $15.27(1)$ \cite{percival1993}  & $33.23(5)$ \\
B0540--69        & $0.051$   & $3.2$  & $49.7$ & $38.69$ &  $22.0(2)$ \cite{Middleditch1987}  & $33.47(15)$ \\
B0833--45 (Vela) & $0.089$   & $4.1$  & $0.28$ & $36.83$ &  $23.7(3)$ \cite{Nasuti1997}  & $28.3(3)$ \\
B0656+14         & $0.385$   & $5.0$  & $0.29$ & $34.58$ &  $25.15(13)$ \cite{Koptsevich2001}  & $27.7^{+0.4}_{-0.9}$\\
B0633+17 (Geminga) & $0.237$ & $5.53$ & $0.25$ & $34.50$ &  $25.7(3)$ \cite{Bignami1993}& $27.37^{+0.16}_{-0.10}$\\
J1023+0038       &  $1.69\times10^{-3}$ & $9.7$ & $1.37$ & $34.63$ & $22.8^{+0.5}_{-0.8}$ $^{[c]}$& $30.03^{+0.33}_{-0.18}$\\
J0337+1715       & $2.73\times10^{-3}$ & $9.4$ & $1.3$ & $34.53$ & $>24.5$ \cite{strader2016} $^{g}$ & $<29.15$ \\
\hline
\multicolumn{7}{c}{\textbf{Non-pulsed optical counterparts of  pulsars}} \\

\hline

J2124-3358 & $4.9\times10^{-3}$ & $9.6$ & $0.41$ & $33.83$ & $27.53^{+0.07}_{-0.06}$ \cite{rangelov2017} $^{g'}$  & $27.14(3)$ \\
B1929+10 & $0.226$ & $6.5$ & $0.31$ & $33.59$ & $\geq 26.2$ \cite{pavlov1996} & $27.26^{+0.2}_{-0.3}$\cite{zharikov2002} \\

B0950+08 & $0.253$ & $7.2$ & $0.26$ & $32.75$ & $27.07(16)$ \cite{zharikov2002} & $26.88(8)$ \\
J0108-1431 & $0.807$ & $8.2$ & $0.21$ & $30.76$ & $27.9(1)$ \cite{mignani2008} & $26.4(2)$ \\
B1133+16 & $1.188$ & $6.7$ & $0.35$ & $31.94$ & $28.1(3)$ \cite{zharikov2013} & $26.76(17)$ \\
J1741-2054 & $0.414$ & $5.6$ & $0.30$ & $33.98$ & $26.45(10)$ & $27.60(5)$ \\
J0205+6449 & $0.066$ & $3.7$ & $3.2$ & $37.43$ & $27.4(1)$ \cite{moran2013} $^{g'}$ & $30.06(10)$ \\
B1509-58   & $0.151$ & $3.2$ & $4.4$ & $37.23$ & $25.7(1)$ \cite{wagner2000,moran2013} $^{R}$ & $31.0(1)$ \\
B1055-52   & $0.197$ & $5.7$ & $0.1$ & $34.48$ & $25.4(1)$ \cite{mignani2010} $^{V}$ & $28.2(2)$ \\

\hline
\end{tabular}
\begin{tablenotes}
\scriptsize \item[$a$] Values taken from ATNF pulsar
catalog\cite{manchester2005}. \item[$b$] magnitudes are de-reddened
and refer to the B band ($\lambda_{eff}=445$~nm,
$\Delta\lambda=66$~nm), unless otherwise indicated by the
superscript; \item[$c$] The equivalent pulsed magnitude of {\1023} in
the B-band was evaluated as $<A>=3.8\times10^{-3}$ times the
deredenned B magnitude of the source averaged over an orbital cycle,
$16.77(5)$~mag\cite{bogdanov2015}. The uncertainty reflects the
standard deviation of $\sigma_A=2.1\times10^{-3}$ of the distribution
of the amplitudes observed over 1.1~ks intervals (see Methods and
Fig.~2 of the the
  main manuscript).
\end{tablenotes}

\end{threeparttable}
\label{tab:S1}
\end{table}

\newpage

\begin{table}
  \centering
  \small
\begin{threeparttable}[h!]
\renewcommand{\tablename}{Supplementary Table}
  
\caption{Log of the  observations of {\1023}.}
\begin{tabular}{lccc}
\hline
     \multicolumn{3}{c}{\textbf{Telescopio Nazionale Galileo (TNG)}}\\
   \hline
 Seq. & Start Time (MJD) & Exposure (s) & Average total countrate ($s^{-1}$)\\
 1 & 57449.9028371 & 3300 & 15732\\
 2 & 57449.9504008 & 3300 & 15231\\
 3 & 57449.9942661 & 3300 & 18180\\
 4 & 57450.0668420 & 3300 & 16837\\
\hline
     \multicolumn{3}{c}{\textbf{Swift XRT} (Id. 33012)}\\
     \hline
     112 & 57446.1256713 &  330.9 & 0.27 \\
     113 & 57450.9222106 &  200.6 & 0.34 \\
     
 \hline
 \end{tabular}
\end{threeparttable}
\label{tab:S2}
\end{table}

\clearpage 
\subsection{Methods reference list.}

\end{document}